%% file: skillos.tex
\documentclass[sigplan, 10pt, nonacm]{acmart}
\renewcommand\footnotetextcopyrightpermission[1]{}
\usepackage{booktabs}
\usepackage{bbding}
\usepackage{tikz}
\usepackage{amsmath}
\usepackage[normalem]{ulem}
\usepackage{graphicx}
\usepackage{pifont}
\usepackage{balance}
\usepackage{multirow}
\usepackage{enumitem}
\usepackage{xspace}
\usepackage[]{hyperref}
\usepackage{algorithm}
\usepackage{algpseudocode}
\usepackage{listings}
\usepackage{xcolor}
\usepackage{makecell}
\usepackage[most]{tcolorbox}

\newtcolorbox{examplebox}{
  colback=gray!5,
  colframe=gray!50,
  boxrule=0.5pt,
  left=6pt, right=6pt, top=4pt, bottom=4pt,
  sharp corners,
  before skip=6pt, after skip=6pt,
}

\usepackage{titlesec}
\titlespacing*{\section}{0pt}{*0.9}{*0.9}
\titlespacing*{\subsection}{0pt}{*0.9}{*0.9}
\titlespacing*{\subsubsection}{0pt}{*0.9}{*0.9}
\begin{document}

\date{}

\title{SkVM: Revisiting Language VM for Skills across Heterogenous LLMs and Harnesses}
\author{Le Chen}
\affiliation{%
  \institution{Shanghai Jiao Tong University}
  \city{Shanghai}
  \country{China}
}

\author{Erhu Feng}
\affiliation{%
  \institution{Shanghai Jiao Tong University}
  \city{Shanghai}
  \country{China}
}

\author{Yubin Xia}
\affiliation{%
  \institution{Shanghai Jiao Tong University}
  \city{Shanghai}
  \country{China}
}

\author{Haibo Chen}
\affiliation{%
  \institution{Shanghai Jiao Tong University}
  \city{Shanghai}
  \country{China}
}

\newcommand{\myparagraph}[1]{\vspace {3pt}\noindent\textbf{\emph{#1}}}

\newcommand{\sys}{\textsc{SkVM}\xspace}

\newcommand{\heading}[1]{
\vspace{1ex}
\noindent
\textbf{#1}
}

\newenvironment{myitemize}%
  {\begin{list}{\labelitemi}{\itemsep1pt \topsep2pt \parsep0.00in
  \partopsep=0pt \leftmargin1em}}%
  {\end{list}}


\input{./abs}

\settopmatter{printfolios=true}
\maketitle

\input{./intro}
\input{./insight}
\input{./overview}
\input{./compiler}
\input{./runtime}

\input{./eval}

\input{./discussion}
\input{./concl}

\bibliographystyle{plain}
\bibliography{references}

\end{document}

%% file: abs.tex
\begin{abstract}
LLM agents increasingly adopt skills as a reusable unit of composition.
While skills are shared across diverse agent platforms, current systems treat them as raw context, causing the same skill to behave inconsistently for different agents.
This fragility undermines skill portability and execution efficiency.

To address this challenge, we analyze ~118,000 skills and draw inspiration from traditional compiler design.
We treat skills as code and LLMs as heterogeneous processors.
To make portability actionable, we decompose a skill's requirements into a set of primitive capabilities,
and measure how well each model-harness pair supports them.
Based on these capability profiles, we propose \sys, a compilation and runtime system designed for portable and efficient skill execution.
At compile time, \sys performs capability-based compilation, environment binding, and concurrency extraction.
At runtime, \sys applies JIT code solidification and adaptive recompilation for performance optimization.

We evaluate \sys across eight LLMs of varying scales and three agent harnesses, covering SkillsBench and representative skill tasks.
Results demonstrate that \sys significantly improves task completion rates across different models and environments while reducing token consumption by up to 40\%.
In terms of performance, \sys achieves up to 3.2$\times$ speedup with enhanced parallelism, and 19--50$\times$ latency
reduction through code solidification.
\end{abstract}

%% file: intro.tex
\section{Introduction}
\label{sec:intro}

LLM-based agents~\cite{achiam2023gpt4,openclaw,opencode,claudecode,openai-codex-cli,wang2024agentsurvey} are reshaping how software tasks get done.
Instead of writing code by hand, a developer describes a goal,
and the agent reasons, invokes tools, and delivers the result~\cite{yao2023react,qin2023toolllm}.
To support this paradigm, a growing ecosystem of \emph{skills}~\cite{claude-agent-skills,anthropic-skills-repo,openclaw-medical-skills,skillssh} has emerged
to make agent execution more reliable and practical.
Skills primarily consist of natural language descriptions and scripts, encapsulating domain-specific
procedures and best practices~\cite{anthropic-skills-blog}.
Over 100,000 skills are now distributed across major platforms~\cite{clawhub, skillssh},
spanning data analysis, finance, office automation, programming,
and beyond.

However, current agents' support for skills is simplistic: skills are treated as
additional context and passed directly to the model. Models differ substantially in
their ability to understand and execute skills~\cite{lu2024agentbench, han2026swe, li2026skillsbench}.
Across eight models spanning a range of scales,
enabling skills degrades performance on 15\% of tasks overall (e.g., 7\% for Opus 4.6 and 25\% for Qwen3-30B).
For another 17\% of tasks, scores remain unchanged (excluding tasks with 100\% completion),
and on up to 87\% of tasks, at least one model shows no improvement after skill usage.
Recent work~\cite{han2026swe} has reached similar conclusions, finding that on the SWE-Benchmark, 39 out of 49 skills showed no score improvement, and 3 skills experienced significant score degradation.

\begin{figure}[t]
  \centering
  \setlength{\abovecaptionskip}{5pt}
  \setlength{\belowcaptionskip}{-10pt}
  \includegraphics[width=\columnwidth]{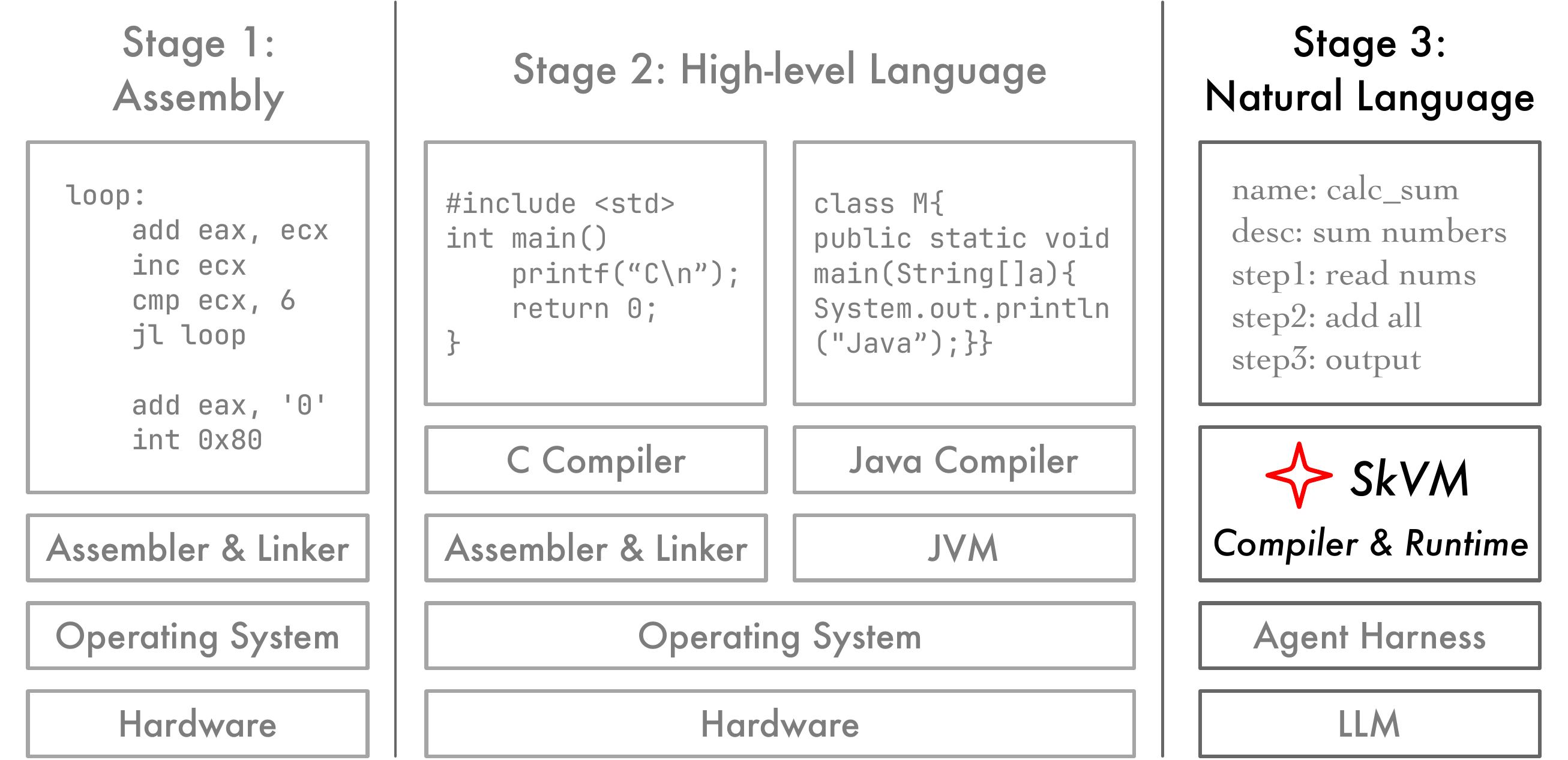}
  \caption{Evolution of programming abstractions. Skills are
    the current frontier: natural language programs of hundreds of lines
    that lack a compiler and runtime for cross-target portability.}
  \label{fig:evolution}
\end{figure}

We attribute these issues to two factors: (1) even when a skill is loaded, the model may ignore its guidance during execution;
and (2) even when the model follows the skill, the skill's assumptions about model capabilities may not match the model's actual competence.
Beyond differences in task completion gains, the semi-structured code snippets and workflow-style formats
common in current skills also incur substantial token overhead (e.g., 451\% token increase while pass rates remain unchanged~\cite{han2026swe}) and higher latency during execution.

This problem arises from a fundamental mismatch between static skills and the variability of underlying models and agent harnesses.
The capabilities a skill demands may not align with the capabilities
of the LLM invoked at runtime. Revisiting the evolution of computing paradigms~\cite{stroustrup2013c++, venners1998java, cramer1997compiling} (as shown in Figure~\ref{fig:evolution}), we observe that
in the agent era, \textbf{skills are code}, and \textbf{LLMs are processors}. Yet today, no mechanism
exists to efficiently and reliably execute skills across heterogeneous LLMs and agent harnesses.

Inspired by how classical computing systems handle code, we introduce
\sys, a compilation and runtime system for skills that enables
efficient cross-model execution and provides unified underlying
support for skill execution. Through \sys, skills are compiled
into forms tailored to different models, allowing each model to
better understand and execute skills. Furthermore, \sys provides a unified runtime
environment that systematically manages skill loading, parsing,
and concurrent execution.

\textit{We structure \sys around classical compilation techniques:
interpreted execution~\cite{jansen2007interpretation}, ahead-of-time (AOT) compilation~\cite{serrano2021javascript, serrano2018javascript}, and just-in-time (JIT)
optimization~\cite{aho2006compilers, aycock2003jit, suganuma2000overview, ishizaki2000study, cramer1997compiling}.} Currently, agents handle skills using only interpreted
execution, feeding raw skill text directly to the model. This approach
hampers skill portability and execution efficiency. In contrast, \sys further applies
AOT and JIT compilation to optimize for diverse backend models and execution
environments:

\textbf{AOT compilation:} Skills inherently encode requirements of model capabilities, environment configuration, and parallelism opportunities. 
Therefore, \sys analyzes these requirements before execution and optimizes skills accordingly.
\emph{Capability-based compilation} characterizes the capability gap between LLMs and skill requirements by extracting 26 primitive capabilities. 
Each capability abstracts a distinct dimension of model behavior with multiple proficiency levels.
The compiler measures the target model against these capability dimensions and adapts the skill specification to better align with model strengths and limitations.
Second, \emph{environment binding} extracts implicit dependencies on packages and tools from skill descriptions and generates setup scripts executable at load time, ensuring all dependencies are satisfied before skill execution.
Finally, \emph{concurrency extraction} draws inspiration from classical compiler optimizations for data-level parallelism (DLP)~\cite{kalathingal2016dynamic}, instruction-level parallelism (ILP)~\cite{ranganathan1998empirical, kastner2001ilp}, and thread-level parallelism (TLP)~\cite{redstone2003mini}. It extracts parallelism opportunities at similar three granularities from skills and explicitly exposes them to the agent harness, thereby improving task execution efficiency.

\textbf{JIT optimization:} At runtime, \sys leverages runtime information to continuously optimize skill performance and correctness. 
\emph{Code solidification:} Skills may contain parameterized script templates which may be instantiated multiple times during different executions. 
JIT compilation materializes these high-frequency templates into instantiated executable code, bypassing LLM parsing. 
\emph{Adaptive recompilation:} JIT monitoring tracks model execution and adaptively recompiles skills when
capability gaps emerge, ensuring skills can self-improve through iterative optimization.

After skill compilation completes, the skill runtime parses compiled skill artifacts (including optimized
skills, instantiated scripts, and concurrency dependency graphs) rather than
simply passing raw text to the model. Additionally, the skill
runtime coordinates with the system's available resources and tool capabilities
to schedule agent execution in real time, and ensures stable and reliable agent execution.

We evaluate \sys across eight LLMs of varying scales and three agent harnesses, covering SkillsBench~\cite{li2026skillsbench} and representative skill tasks (118 tasks).
Results demonstrate that \sys significantly improves task completion rates (averaging 15.3\%) across different models and agent harnesses~\cite{claudecode,opencode}.
Furthermore, for tasks that can be completed, \sys reduces token consumption up to 40\%.
In terms of performance, \sys achieves 3.2$\times$--50$\times$ wall-clock speedups through fine-grained parallelization and JIT code solidification.

%% file: insight.tex
\section{Skills in the Wild}
\label{sec:insight}

\begin{figure*}[t]
  \centering
  \setlength{\abovecaptionskip}{0pt}
  \setlength{\belowcaptionskip}{-8pt}
  \includegraphics[width=\textwidth]{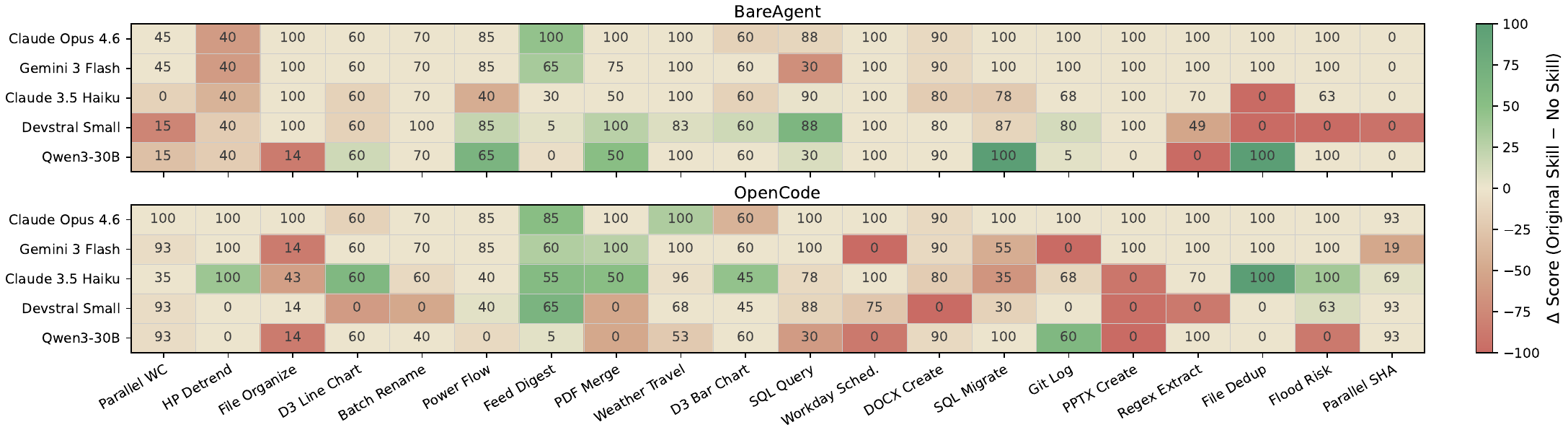}
  \caption{Skill performance across models and harnesses.
    Columns are tasks, rows are models, and each subplot corresponds to one harness.
    Cell colors indicate score deltas relative to the no-skill baseline,
    and cell labels show absolute task scores.
    Both model identity and harness choice significantly affect outcomes.}
  \label{fig:m2_heatmap}
\end{figure*}

In this section, we first introduce skills, the agent that uses them,
and how skills are invoked in practice.
We study the skill ecosystem at scale through more than 110,000 skills
collected from two major distribution platforms:
clawhub.ai~\cite{clawhub} (28,990 skills) and skills.sh~\cite{skillssh} (89,280 skills).
Based on this ecosystem analysis, we further conduct experiments to
identify concrete problems in current skill usage.

\subsection{Skill Definition}
\label{sec:back:skills}

A \emph{skill} is a distributable, self-contained knowledge pack that
enhances an AI agent's capability on a specific class of
tasks~\cite{claude-agent-skills, agentskills-spec}.
A skill may introduce knowledge the base model does not have, or
constrain the model toward a prescribed workflow when it already
has the relevant capability but would otherwise improvise.
For example, a PDF-processing skill~\cite{anthropic-skills-repo} teaches the model how to use pdfplumber for table
extraction, and also constrains the model to use pypdf instead of the deprecated PyPDF2 when merging PDF files.

A skill is typically defined in a structured file comprising three
layers:
(1)~\emph{metadata}---a name, description, and trigger conditions
that allow the agent runtime to discover and select the skill automatically;
(2)~\emph{instructions}---the main body of a skill that encodes
multi-step workflows, tool references, and other task-specific guidance
in natural language; and
(3)~\emph{bundled resources}---scripts, references, and
templates stored in companion files.

This structure distinguishes skills from ad-hoc prompt snippets.
The metadata makes a skill a discoverable, selectable unit, enabling the
agent runtime to match tasks to applicable skills automatically rather than
requiring manual prompt assembly~\cite{claude-code-skills}.
Compared with fine-tuning, skills require no weight modification
and can be operated entirely at the prompt level~\cite{ouyang2022training, liu2021pretrain}, making them lightweight to
distribute and easy to compose.
Multiple agent platforms have converged on this
abstraction~\cite{claude-code-skills, cursor-skills, openai-codex-cli, agentskills-spec, antigravity-skills, coze-skills}, and existing
tools support basic skill authoring, management, and optimization.

\subsection{LLM Agent and Agent Harness}
\label{sec:back:harness}

An LLM agent works in a ReAct loop~\cite{yao2023react}. It receives a task, reasons
about what to do, issues a tool call (reading a file, running
a command, making an API request)~\cite{schick2023toolformer, openai2023functioncalling}, observes the result, and
reasons again. This loop repeats until the task is done.

The agent does not interact with the outside world directly.
It runs inside an \emph{agent harness}, a runtime framework
that sits between the model and the operating environment~\cite{wu2023autogen, langgraph}.
The harness manages the LLM API connection, registers the
tools the agent can call, dispatches those calls, and controls
how much conversation history fits in the context window.
Claude Code~\cite{claudecode}, OpenCode~\cite{opencode}, and OpenClaw~\cite{openclaw} are examples of widely used
harnesses.

Harnesses differ in ways that matter for skills~\cite{google-a2a}.
Some provide a rich set of file-system and web search tools,
while others expose only basic read, write, and shell execution operations.
Some can spawn independent sub-agents that work concurrently,
while others only support sequential execution.
These differences directly affect whether a skill's instructions can be carried out.

\subsection{Characterizing Skills}
\label{sec:insight:analysis}

\begin{figure}[t]
  \centering
  \setlength{\abovecaptionskip}{0pt}
  \setlength{\belowcaptionskip}{-10pt}
  \includegraphics[width=\columnwidth]{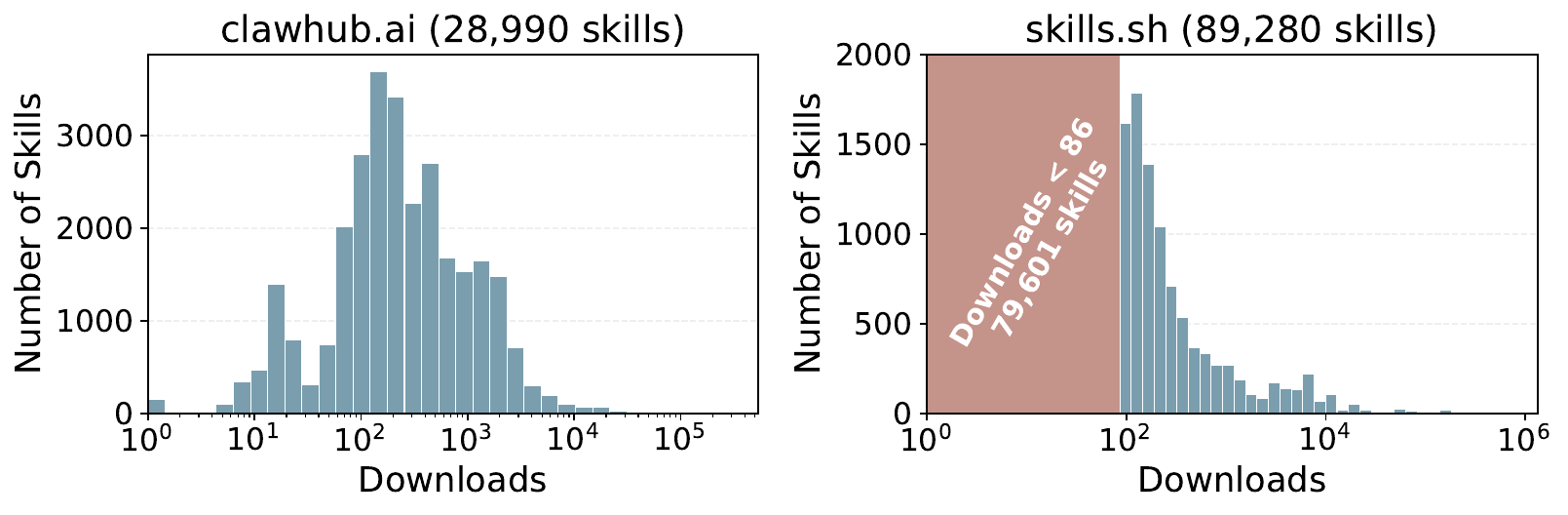}
  \caption{Skill download distribution on clawhub.ai and
    skills.sh. Both platforms show a long-tailed distribution.}
  \label{fig:skill_distribution}
\end{figure}

\heading{Ecosystem Scale and Distribution.}
The skill ecosystem is large but uneven.
Figure~\ref{fig:skill_distribution} shows the download counts
across both platforms. On skills.sh, 89\% of the 89,280 skills
have fewer than 86 downloads, while a handful of head skills
reach $10^4$--$10^6$. clawhub.ai follows a similar long-tailed
pattern, peaking around $10^2$--$10^3$ downloads. In total,
over 118,000 skills exist across the two platforms, but the
vast majority see little use.

\heading{Skill Taxonomy.}
Skills differ substantially in what they provide the agent with.
We use LLM-assisted analysis to classify 15,063 skills that have
more than 100 downloads into three categories, based on which
aspect the skill most emphasizes: tool reference (52\%),
procedural guidance (28\%), and content generation (20\%).

\emph{Tool-reference skills} teach the model how to
operate specific tools, APIs, or CLIs. They are essentially
usage documentation loaded into the agent's context.
A representative example is a skill that documents
how to use python to merge, split, and manipulate PDF files.

\emph{Procedural skills} prescribe step-by-step
workflows and reasoning strategies for carrying out a task.
A representative example is a debugging skill that enforces
a four-phase process: reproduce the failure, trace the root
cause, apply a targeted fix, and verify the fix with tests.

\emph{Generative skills} ask the model to produce
content, code, documents, or prose. The core quality depends
on the model's own capability. An example is a
frontend-design skill~\cite{anthropic-skills-repo} that generates production-grade web
components with specific design guidelines and styling
conventions.

Tool-reference and procedural skills together account for
80\% of the ecosystem. These skills encode analyzable
structure, including script fragments and step-by-step workflows.
The correctness of executing these skills depends
on whether the agent executes the prescribed steps, not on
its generative ability.

\heading{Workflow Structure.}
Beyond task type, skills also exhibit repeated internal
structure. Across the 15,063 classified skills, 76\%
contain explicit procedural structure: numbered steps,
conditional branches, and data dependencies between steps.

\heading{Code Fragments.}
75\% of skills embed code-like fragments: shell
commands, API call patterns, or script snippets whose overall
shape recurs across invocations while only the input-specific
parameters change.
For example, a weather skill~\cite{openclaw} may provide a family of curl command
templates for current conditions, forecasts, and format options, where
the command structure stays fixed while parameters such as the city,
output format, and forecast day vary.

\subsection{Problems with Current Skill Usage}
\label{sec:insight:problems}

Skills are loaded into the agent's context as raw text, with no adaptation.
But skills carry implicit assumptions about the model, agent harness, and user environment.
When those assumptions are wrong, three failure modes emerge.

\heading{P1: Model Mismatch.}
A skill implicitly assumes that the model is capable enough
to follow its instructions. In practice, models differ
dramatically~\cite{lu2024agentbench}. 
Figure~\ref{fig:m2_heatmap} shows scores for
a subset of tasks across three models and three harnesses.\footnote{BareAgent is a minimal agent harness that we built. It implements only basic agent runtime logic and exposes only read, write, and exec tools.}
Performance varies substantially across models on the same skill.
Moreover, enabling skills degrades performance on 15\% of tasks; leaves scores unchanged on 17\% of tasks (excluding tasks with a 100\% completion rate); and yields no improvement for at least one model on 87\% of tasks.
\begin{examplebox}
\textbf{Case Study:} the pptx-creation skill~\cite{anthropic-skills-repo}
recommends PptxGenJS (a javascript library) for slide generation.
Claude-opus-4.6 and gemini-3-flash both score 100 with this skill,
but devstral-small misreads PptxGenJS as a CLI when using OpenCode
and repeatedly executes wrong commands.
Without the skill, devstral-small instead uses python-pptx and scores 95.
The skill assumes a model can distinguish a library API from a CLI,
which holds for frontier models but not weaker ones.
\end{examplebox}

\heading{P2: Harness Mismatch.}
The same model produces different results on the same task depending on the agent harness it runs in~\cite{yang2024sweagent}.
Figure~\ref{fig:m2_heatmap} also shows this dimension:
across the heatmap, harness-induced variance is comparable to model-induced variance.
Skills are written without knowledge of what tools and plugins the harness provides.

\begin{examplebox}
\textbf{Case Study:} Gemini 3 Flash scores 100 on the workday-scheduling task
with the original skill on BareAgent, but 0 on OpenCode with the same skill.
The failure is triggered by a malformed JSON key that makes the output
unparseable. BareAgent adds only a minimal system prompt, whereas
OpenCode prepends extensive tool documentation, and the longer combined
context induces the formatting error.
\end{examplebox}

\begin{figure}[t]
  \centering
  \setlength{\abovecaptionskip}{3pt}
  \setlength{\belowcaptionskip}{0pt}
  \includegraphics[width=\columnwidth]{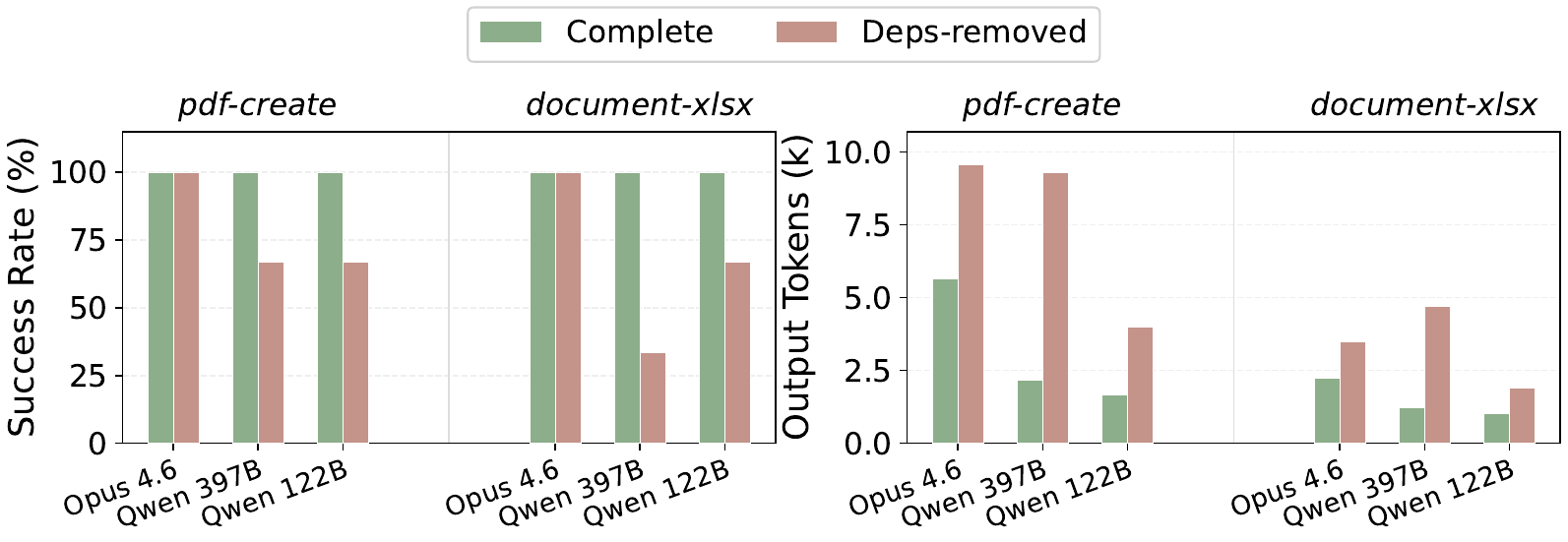}
  \caption{Removing a required dependency hurts both correctness and efficiency.}
  \label{fig:p3_env_motivation}
\end{figure}

\heading{P3: Environment Mismatch.}
Skills have dependencies on the execution environment, but
the user's machine may lack the required packages, tools, or
configurations.
Some skills list prerequisites, but those instructions are often general and target
a generic machine rather than the user's actual setup. The
real requirement depends on the current OS, hardware, installed
versions, and package-manager state.
Other skills omit the dependency entirely.
Figure~\ref{fig:p3_env_motivation} shows the effect on two
representative tasks. When the required library is absent,
both Qwen models drop to 33--67\% success rate while
generating 2--4$\times$ more output tokens on failed
workarounds. Even Claude~Opus~4.6, which still succeeds on
every run, generates 56--69\% more output tokens because it
must diagnose the mismatch and install the missing package
at runtime. Leaving diagnosis to the model therefore turns
every missing dependency into a repeated tax on both
correctness and efficiency.

\subsection{Challenges}
\label{sec:insight:challenges}

Skills are distributed as plain Markdown so that any agent can load them with no special tooling.
The mismatches above undermine this portability promise.
Restoring portability requires adapting each skill to its
target, which is challenging for two reasons.

\heading{C1: Targets Differ Along Many Axes.}
Each of the three dimensions above, model, harness, and
environment, varies independently. A skill that needs
adaptation along one dimension may need a different adaptation
when two dimensions change together. The right fix for a weak
model on a rich harness differs from the right fix for the
same model on a minimal one. This interaction makes the
adaptation space combinatorial: it cannot be covered by a
fixed set of rewrite rules.

\heading{C2: Skills Are Unstructured Natural Language.}
Source code requires compilation before it can run on a target system,
and has a grammar, types, and a well-defined AST that tools can analyze~\cite{aho2006compilers}.
In comparison, skills are written in natural language.
Although skills are mostly created following common conventions,
different skills vary widely in structure, terminology, and level of detail.
Before any adaptation can happen, the skill's implicit requirements and workflow structure
must be recovered from prose.

%% file: overview.tex
\section{Design Overview}
\label{sec:overview}

\begin{figure}[t]
  \centering
  \setlength{\belowcaptionskip}{-10pt}
  \includegraphics[width=\columnwidth]{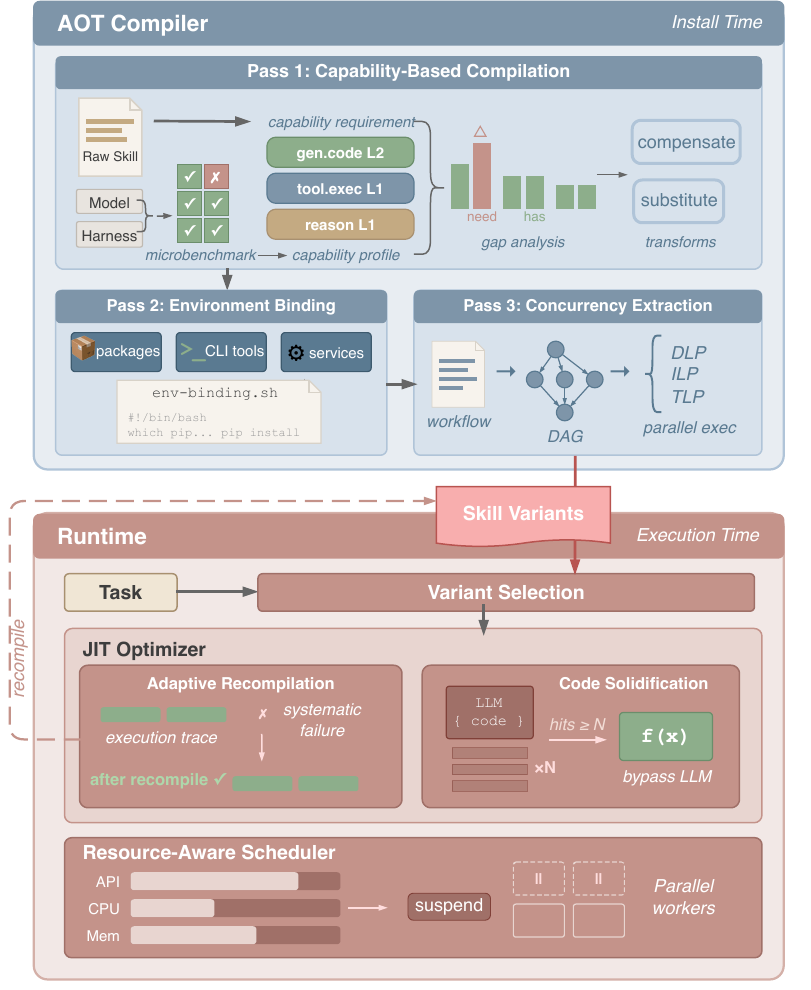}
  \caption{\sys architecture. The AOT compiler produces
    optimized skill variants at install time through three
    passes. The runtime manages variant selection, JIT
    optimization, and resource-aware scheduling during
    execution.}
  \label{fig:overview}
\end{figure}

\sys is a compilation and runtime system for skills.
It makes skills portable across heterogeneous targets by
combining an AOT compiler~\cite{aho2006compilers} that specializes skills at install time
with a runtime that resolves execution-time uncertainty.
Figure~\ref{fig:overview} shows the architecture.

\heading{AOT Compiler.}
At skill installation time, the compiler analyzes the
skill against the target environment and produces one or more optimized
\emph{skill variants}. The compiler performs three distinct passes:
\emph{Capability-based compilation}
(\S\ref{sec:compiler:cap}) handles model and harness
mismatch (P1--P2) by adapting the skill to the target's
available capabilities.
\emph{Environment binding}
(\S\ref{sec:compiler:env}) handles environment mismatch
(P3) by materializing the dependencies required for
execution.
\emph{Concurrency extraction}
(\S\ref{sec:compiler:par}) identifies parallelism latent in
the skill's workflow and maps it to the target harness.
Compiled variants are stored per target.

\heading{Runtime \& JIT optimizer.}
When a task arrives, the runtime selects the variant
compiled for the current (model, harness) pair and loads
it through the standard progressive disclosure mechanism.
Two components then operate on top of normal execution.
The \emph{JIT optimizer} (\S\ref{sec:runtime:jit})
monitors execution outcomes across invocations and
triggers recompilation when it detects capability gaps
that the AOT compiler missed.
It also compiles structurally fixed code patterns into
executable functions that bypass the LLM inference entirely.
The \emph{resource-aware scheduler}
(\S\ref{sec:runtime:sched}) bridges compile-time
parallelism annotations and run-time resource availability,
throttling or suspending concurrent sub-agents when
demand exceeds capacity.

%% file: compiler.tex
\section{Skill Compilation}
\label{sec:compiler}

When a user installs a skill for the first time, \sys compiles it against the target (model, harness, host environment) before the skill ever runs.
The compiler reads the raw skill text, identifies the target,
and produces one or more optimized \emph{skill variants} through three sequential passes.
Each pass closes a different kind of gap between what the skill expects and what the target provides.

\subsection{Capability-Based Compilation}
\label{sec:compiler:cap}

The first pass addresses the model and harness mismatch.
Because targets vary widely (\S\ref{sec:insight:challenges}),
the compiler cannot embed optimization logic specific to each target.
Instead, \sys introduces \emph{primitive capabilities},
an abstract vocabulary for expressing what a skill requires and what a target can provide.
The compiler profiles each target offline with microbenchmarks for these capabilities,
compares the profiling result with the skill's capability requirements,
and selects the optimization strategy based on the gap.

\subsubsection{Primitive Capabilities}

Just as JVM bytecodes define the basic operations a Java program requires from its runtime,
primitive capabilities define the basic abilities a skill requires from its target.
A primitive capability is an indivisible unit that describes what is needed to complete the tasks a skill defines.
It is defined purely from the demand side, independent of any specific model.
The definition of primitive capabilities follows three principles:
\begin{myitemize}
  \item \emph{Composability}: any concrete task within a skill can
    be described as a combination of primitive capabilities.
  \item \emph{Generality}: each primitive capability must be broad
    enough to appear across many skills, keeping the total
    capability set small and the compiler's workload bounded.
  \item \emph{Semantic independence}: a primitive capability concerns structural correctness,
  rather than whether the output is insightful or well-argued.
\end{myitemize}

We derived the primitive capability set from a corpus of 15,063 skills
in two stages. First, we curated 50 representative skills spanning
document generation, script execution, data analysis, and code
development, and used LLM-assisted analysis guided by the three
principles above to decompose them into an initial set of 19
primitive capabilities. We then manually reviewed each primitive capability to verify that
it satisfies all three principles. Second, we validated the initial
set against the full corpus of 15,063 skills by checking whether
each skill's requirements can be expressed as a combination of the
existing primitives. Skills that could not be covered revealed
missing capabilities. A missing capability was added to the set only
when it appeared in at least 1\% of the corpus, filtering out
rare or domain-specific requirements. This process converged on
26 primitive capabilities across four categories that cover the
requirements of 95\% of the skills.

We also observe that different tasks demand the same
primitive capability at different depths. We therefore
define multiple proficiency levels for each primitive capability.
A higher level represents deeper use of the capability: an \emph{L(n+1)} requirement
cannot be satisfied by \emph{L(n)} ability.
Table~\ref{tab:primitives} shows representative examples.
For each primitive capability level, we generate a suite of microbenchmarks
and evaluate them across different models.
If a model succeeds on higher-level microbenchmarks while failing on lower-level ones,
we interpret this inconsistency as evidence that the level hierarchy is miscalibrated.
We then revise the level definitions and repeat the evaluation.

\begin{table}[t]
\setlength{\abovecaptionskip}{4pt}
\setlength{\belowcaptionskip}{0pt}
\centering
\caption{Representative primitive capabilities and their
  proficiency levels. The full catalog contains 26 primitive capabilities
  across four categories.}
\label{tab:primitives}
\resizebox{\columnwidth}{!}{%
\begin{tabular}{@{}llll@{}}
\toprule
\textbf{Primitive} &
  \textbf{L1} & \textbf{L2} & \textbf{L3} \\
\midrule
gen.code.shell &
  Basic commands (ls, cat) &
  Pipes, redirection, loops &
  Complex pipelines (sed, awk) \\
\addlinespace
reason.arithmetic &
  Single-step ops &
  Multi-step &
  Compound \\
\addlinespace
tool.exec &
  Single command &
  Params and relative paths &
  Chained multi-step execution \\
\addlinespace
follow.procedure &
  3 sequential steps &
  5--7 with branches &
  Loops and verification \\
\bottomrule
\end{tabular}%
}
\end{table}

\subsubsection{Measuring the Gap Between Skill and Target}

Before applying optimizations, the compiler must determine
the gap between what the skill requires and what the target
provides. On the skill side, the compiler extracts the
skill's capability requirements at install time: it
decomposes the skill into concrete tasks and maps each task
to the primitive capabilities it requires. 

On the target side, \sys profiles the target with
microbenchmarks generated for each primitive capability to
measure how well the target supports it.
Success on a microbenchmark for a primitive capability at a given level
indicates that the model has attained at least that level of proficiency for that capability.
This profiling is performed when the user sets up the agent and model.
The results are cached and reused for all skill compilations
for the same target.

\subsubsection{Capability-Aware Skill Transforms}

With the skill's capability requirements and target's capability profile in hand, the compiler
walks through each capability requirement and selects a
transform based on the gap between the required level and
the target's profiled level.

\textbf{\emph{Compensation}} applies when the target supports the
required capability but at a lower level.
The compiler first identifies the distinctions between the two levels of a primitive capability.
It then transforms the skill accordingly to reduce its capability requirement from a higher level to one supported by the target model.
Such transformations may include providing examples, making instructions more explicit, and strengthening task constraints.
Compensation is the preferred transform because it preserves the skill's original intent.

\textbf{\emph{Substitution}} is the fallback when compensation is
not viable: the target lacks the capability entirely, or the
level gap is too large for compensation to bridge reliably.
The compiler switches to an alternative
implementation path that achieves the same goal using
different capabilities. For example, if a skill requires
gen.code.python at L3 for a pandas workflow but the target
only reaches L1, the compiler can switch to an SQL-based
path if the target's gen.code.sql reaches L2. These
alternative paths form \emph{equivalence classes} that the
compiler identifies during requirement extraction.

\begin{figure}[t]
  \centering
  \setlength{\abovecaptionskip}{5pt}
  \setlength{\belowcaptionskip}{0pt}
  \includegraphics[width=\columnwidth]{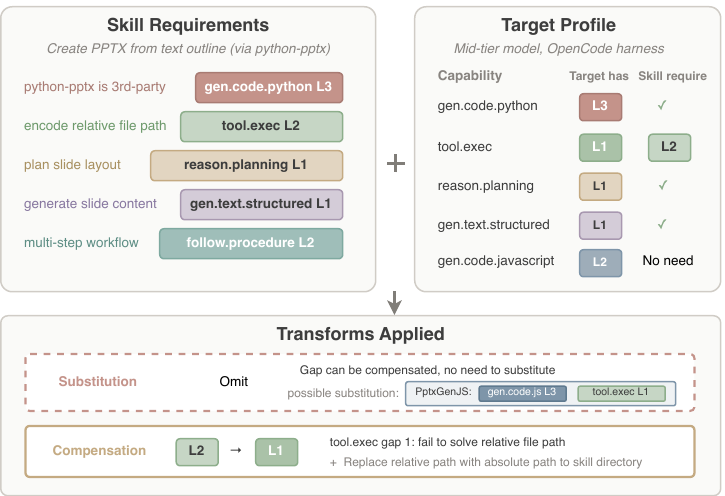}
  \caption{Capability-based compilation on a PPTX skill.
    The compiler extracts skill requirements (left),
    profiles the target (middle), and applies transforms
    based on the gap (bottom).}
  \label{fig:compilation_case}
\end{figure}

\heading{Case Study.}
Figure~\ref{fig:compilation_case} illustrates the
compilation process on a real skill that teaches an agent
to create PPTX presentations using python-pptx.
The skill requires five primitive capabilities at varying levels.
The target matches four of them directly but provides
tool.exec at only L1, while the skill requires L2 for
resolving relative file paths.

The compiler identifies a possible substitution path:
switching to \emph{PptxGenJS} via \emph{gen.code.javascript} would replace
python-pptx and lower the tool.exec requirement to L1.
However, because the tool.exec gap is only one level,
compensation is preferred. The L2 profiling results reveal
that this model fails to resolve relative file paths when
executing commands. The compiler injects absolute path
resolution into the compiled skill, replacing relative paths
with absolute paths to the skill directory.

\begin{figure*}[t]
  \centering
  \setlength{\abovecaptionskip}{3pt}
  \setlength{\belowcaptionskip}{0pt}
  \includegraphics[width=\textwidth]{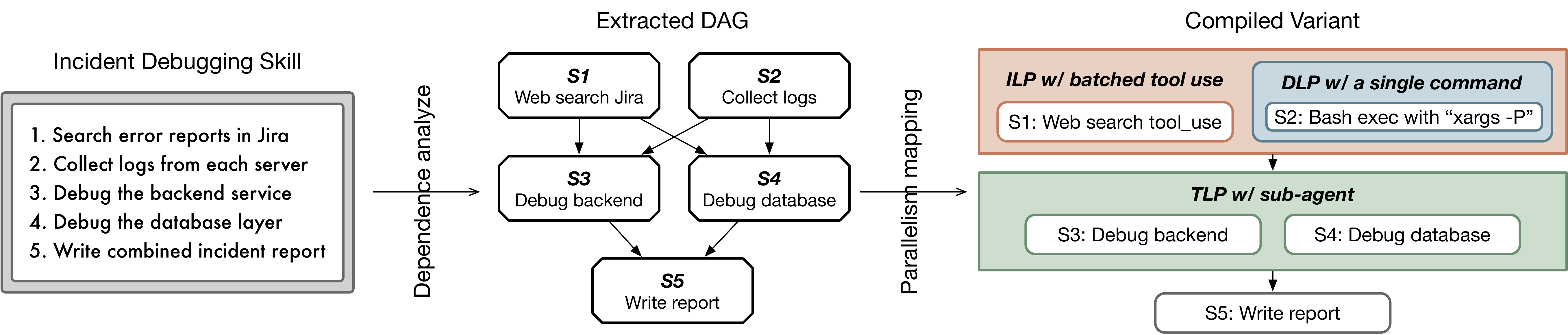}
  \caption{Concurrency extraction on an incident-response skill.
    The compiler decomposes sequential steps into a workflow DAG,
    identifies parallel opportunities both between steps and within
    steps, and maps each to the appropriate concurrency primitive.}
  \label{fig:concurrency_extraction}
\end{figure*}

\subsection{Environment Binding}
\label{sec:compiler:env}

The second pass resolves \emph{P3: environment mismatch}.
Rather than merely checking whether dependencies are installed,
the compiler uses a built-in \emph{env-binding skill} to
reconcile the skill's assumptions against the host environment and
generate an env-binding script.

At skill installation time, the compiler first extracts a dependency manifest from the output
skill of the first pass and any explicit prerequisites, covering external libraries, CLI tools, and system services.
For each manifest entry, the compiler runs a lightweight presence check (e.g., pip show, which).
Dependencies that are already satisfied require no further work.
When the check fails, the compiler performs deeper, system-aware probing to
determine how to resolve the dependency.
The compiler then emits an idempotent env-binding script which will check and repair the dependency before skill execution.
When the environment changes or the env-binding script fails,
execution falls back to the model, and the script's output is passed to the model as additional context.

\subsection{Concurrency Extraction}
\label{sec:compiler:par}

The third pass finds parallelism hiding in plain sight.
Recall that 76\% of skills contain procedural structure
(\S\ref{sec:insight:analysis}). These skills describe
multi-step workflows in sequential prose, but not every step
actually depends on the one before it.

The compiler uses LLM-assisted analysis to decompose the
skill into discrete steps, identifying for each step what
inputs it consumes and what outputs it produces. It then
builds a workflow DAG: nodes are steps, and an edge from
step A to step B exists when B consumes an output of A.
The compiler also analyzes each step internally for
sub-operations that can proceed concurrently.
Figure~\ref{fig:concurrency_extraction} illustrates this
process on an incident-response skill.
For each parallel opportunity, whether between steps or within a
step, the compiler selects a concurrency primitive based
on the task structure and the harness. \sys defines three
levels of parallelism.

\myparagraph{Data-Level Parallelism (DLP).}
A single step applies the same operation to multiple
independent data items, such as running the same analysis
on each of 15 CSV files. The compiler rewrites the step to
process items concurrently using language-level parallelism
primitives appropriate to the skill's implementation, such
as shell-level parallel execution, Python's multiprocessing,
or JavaScript's Promise.all. DLP requires no special
harness support.

\myparagraph{Instruction-Level Parallelism (ILP).}
Multiple independent steps each need a tool call with no
data dependency between them, such as running eight
independent code-analysis scripts on a project. The compiler
groups the corresponding DAG nodes into one parallel stage
and rewrites the skill so that their tool invocations are
issued together in a single LLM turn, instead of one after
another in sequential prose. The compiler annotates the skill with which
step outputs should be bound back to which downstream DAG
inputs after the batch completes. ILP requires the
harness to support batch tool dispatch.

\myparagraph{Thread-Level Parallelism (TLP).}
The workflow decomposes into independent sub-tasks that each
require multi-turn reasoning, such as debugging three
independent services. The compiler rewrites the skill to
extract each sub-task into a separate sub-agent block with
an explicit task description, declared input context, and
expected output that can be consumed by the parent workflow.
The annotation marks this block for sub-agent dispatch, so
the runtime spawns one sub-agent per block in its own
session and merges the returned outputs back into the DAG.
TLP requires harness support for sub-agent spawning.

Opportunities whose required primitive is unavailable on the
target harness fall back to sequential execution.
For each viable opportunity, the compiler either rewrites
the skill's instructions to execute items concurrently (DLP)
or emits execution annotations that make the parallel
structure explicit: a parallel stage over DAG nodes for ILP
or a sub-agent task block with input and output contracts
for TLP. The runtime reads these annotations and dispatches
them with the corresponding harness primitive.

%% file: runtime.tex
\section{Skill Runtime with JIT Optimization}
\label{sec:runtime}

\begin{figure*}[t]
  \centering
  \setlength{\abovecaptionskip}{3pt}
  \setlength{\belowcaptionskip}{0pt}
  \includegraphics[width=\textwidth]{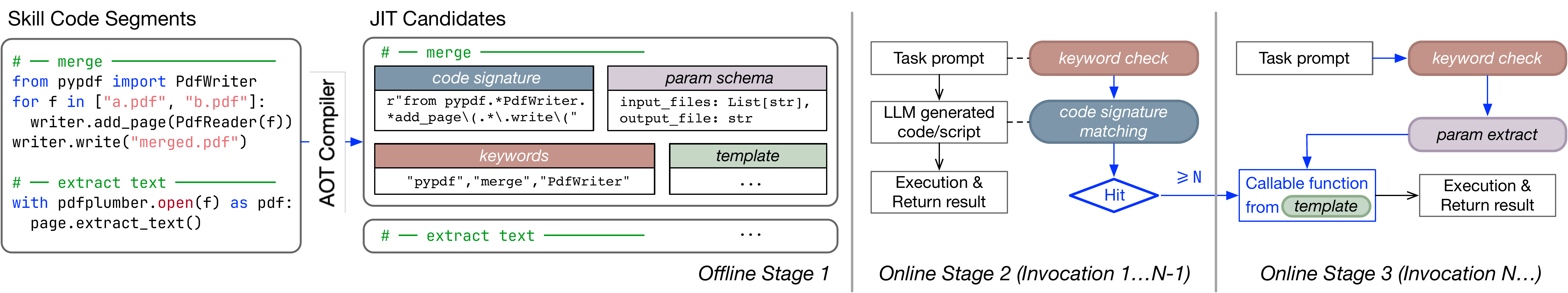}
  \caption{Code solidification pipeline. The AOT compiler
    analyzes skill code segments and generates JIT candidates
    with code signatures and templates (Stage~1). The runtime
    validates predictions across invocations (Stage~2). After
    promotion, a compiled function replaces LLM inference
    (Stage~3).}
  \label{fig:solidification}
\end{figure*}

AOT compilation generates multiple skill variants for different targets.
Consequently, at runtime, \sys loads the skill variant compiled for the current target.
Since AOT compilation is performed only once per skill and target (model+harness) pair, the compilation overhead is modest.

However, some problems only surface at execution time, such as
capability gaps that static profiling missed or resource
contention among parallel sub-agents competing for
rate-limited APIs. Other opportunities, like code patterns
that recur across invocations, only become visible after
repeated runs. The \sys runtime addresses both through
JIT optimization (\S\ref{sec:runtime:jit}) and a
resource-aware scheduler (\S\ref{sec:runtime:sched}).

\subsection{Skill Registration and Execution}
\label{sec:runtime:exec}

Compilation produces multiple skill variants for each skill,
along with the env-binding scripts generated by Pass~2.
When a new task arrives, the runtime selects the variant
compiled for the current (model, harness) pair.
From the agent's perspective, the complexity of multiple
variants is invisible: it sees a single set of skills and
loads them through the same progressive disclosure mechanism
described in \S\ref{sec:insight}. Before loading the full
skill context, the runtime will first execute the
env-binding script to verify whether runtime dependencies are
satisfied. 

\subsection{JIT Optimization}
\label{sec:runtime:jit}

Beyond AOT compilation, which tailors primitive capabilities
to each target, \sys further applies JIT optimization to
enhance execution quality and efficiency through two
complementary mechanisms: adaptive recompilation and code solidification.

\subsubsection{Adaptive Recompilation}
\label{sec:runtime:jit1}

The runtime tracks the outcome of every task executed with
a skill. When a skill execution fails (e.g., agent abort, human feedback) or retries during the agent loop, the runtime records
a structured failure or retry log. Although the current model may perform self-correction through reflection,
the correction process is not recorded, leading to the possibility of encountering the same errors in subsequent executions.

When the same skill fails across multiple invocations, the
runtime analyzes whether the failures are task-specific or reflect a systematic
capability gap in the skill itself. Only in the latter case
does the runtime trigger adaptive recompilation: it feeds
the accumulated failure logs and the model's self-recovery
traces as input to the compiler, which then applies targeted
compensation transforms to the skill.
If the recompiled variant performs worse, the runtime
rolls back to the previous version. Each subsequent
recompilation starts from the best-performing variant
observed so far, ensuring that optimization is improving skill performance.

\begin{figure*}[t]
  \centering
  \setlength{\abovecaptionskip}{0pt}
  \setlength{\belowcaptionskip}{0pt}
  \includegraphics[width=\textwidth]{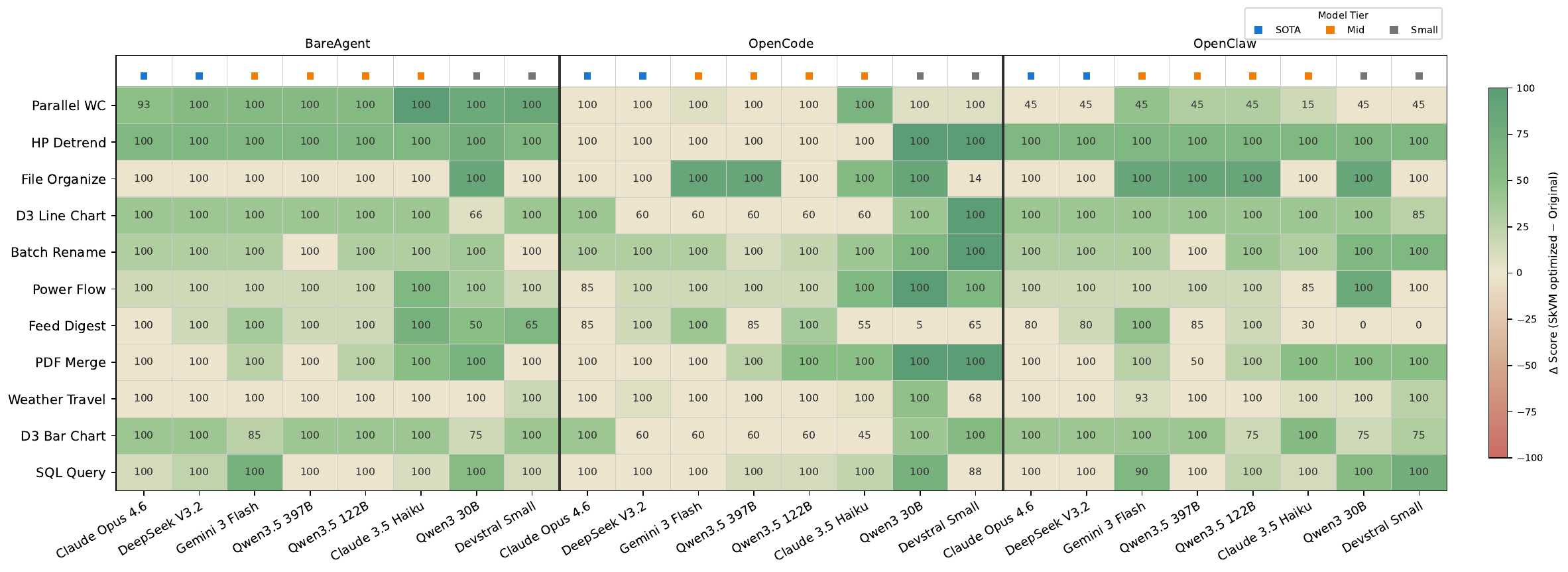}
  \caption{Skill compilation effectiveness. Cell values represent task completion rates after \sys optimization,
  while cell color indicates the score improvement (green) or regression (red) relative to original baseline skills.}
  \label{fig:e1_heatmap}
\end{figure*}

\subsubsection{Code Solidification}
\label{sec:runtime:jit2}
Recall that 75\% of skills contain solidifiable code
(\S\ref{sec:insight:analysis}): code whose overall structure
is fixed, varying only in input-specific parameters.
In normal execution, the LLM re-executes the full
reasoning and tool-use cycle on every invocation, spending
tokens and latency on a process whose structure is identical
each time. Code solidification identifies these
repetitive patterns and promotes them into executable code
that bypasses model inference.
Figure~\ref{fig:solidification} illustrates the three-stage
pipeline.

In the first stage, the AOT compiler analyzes code and
script segments within each skill, identifying whether
they contain parameterized patterns eligible for
solidification. For eligible segments, it generates a
\emph{JIT candidate} containing four components: keywords
that filter relevant invocations, a \emph{code signature}
that captures the expected output structure as a match
pattern, a code template with parameter slots, and a
parameter schema describing the extractable arguments.
These candidates are bundled with the compiled skill variant.

In the second stage, the runtime monitor watches LLM
invocations across repeated calls. For each invocation,
it first checks keywords for relevance, then matches the
generated code against the code signature. 
The code solidification is triggered only after
the code signature matches successfully across multiple consecutive invocations,
ensuring that the code structure generated by the LLM remains consistent with the patterns analyzed in the AOT stage.
If the model's output consistently
diverges from the code signature, the system stays on
the safe LLM path and never promotes.

In the third stage, the template is instantiated into a callable
function: the template code is wrapped with parameter
extraction and output handling to produce a standalone
shell script or code function. This instantiation step
is lightweight and mechanical, as the AOT stage has already
done the analysis. Subsequent invocations bypass the LLM
entirely: the runtime extracts parameters from the task
context, calls the instantiated function, and returns the
result. If the solidified code execution causes an agent
task to fail or encounters a runtime exception, \sys triggers
a fallback mechanism that re-enables LLM-based code generation
to ensure correctness.

\subsection{Resource-Aware Parallel Scheduling}
\label{sec:runtime:sched}

In Pass~3 of AOT compilation, the compiler identifies where parallelism is
possible, but how much parallelism is \emph{efficient}
depends on conditions that change at run time: CPU usage,
external API rate limit, the machine's available
memory, and other shared resources.

\sys dynamically monitors system state and schedules task execution accordingly,
bridging the gap between compile-time opportunity and run-time reality.
For API-bound workloads, it watches response latencies and HTTP 429 (rate limit) signals.
For compute-bound workloads, it monitors CPU and memory usage.
When pressure rises above a threshold, 
\sys applies two mechanisms to mitigate resource contention.
First, it throttles the launch of new sub-agents to avoid introducing additional concurrency pressure.
Second, for sub-agents that are already running, \sys selectively suspends a subset of them (i.e., pauses their agent-loop execution), based on per-agent resource usage or launch order, to reduce active competition for shared resources.
\sys also records the effective concurrency of each skill from its previous run on the current system and uses it as a concurrency hint for subsequent executions.

%% file: eval.tex
\section{Evaluation}
\label{sec:eval}

\begin{figure*}[t]
  \centering
  \setlength{\abovecaptionskip}{0pt}
  \setlength{\belowcaptionskip}{0pt}
  \includegraphics[width=\textwidth]{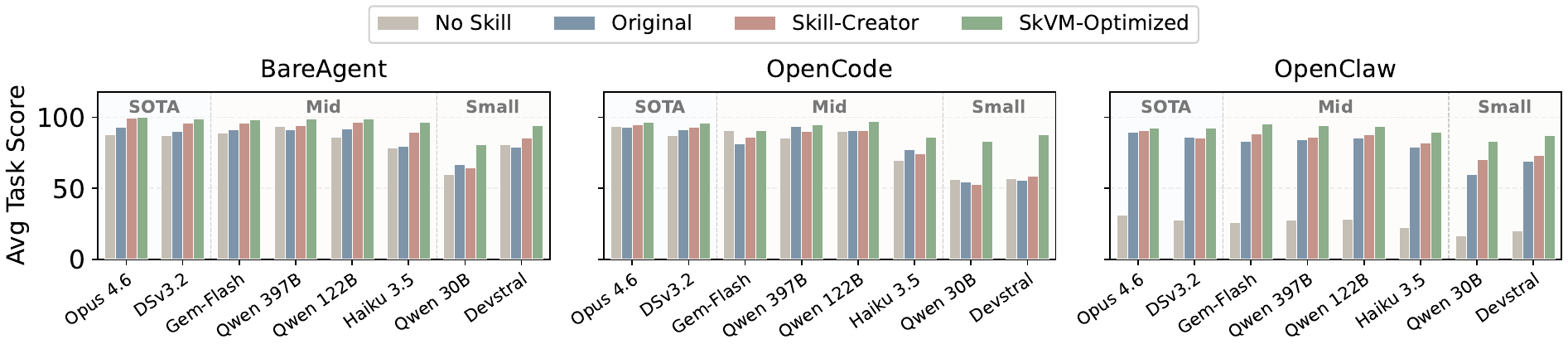}
  \caption{Average task score by skill variant.
    Four variants are compared across eight models and three harnesses:
    No Skill, Original, Skill-Creator, and \sys-Optimized.
    Tier boundaries (SOTA / Mid / Small) are shaded.
    \sys-Optimized skills consistently achieve the highest scores, with the
    largest absolute gains for weaker models and on OpenClaw.}
  \label{fig:e1_overview}
\end{figure*}

\subsection{Setup}
\label{sec:eval:setup}

\heading{Benchmark.}
We select and extend existing skill benchmarks, including SkillsBench~\cite{li2026skillsbench} and PinchBench~\cite{pinchbench}, covering diverse agent tasks spanning code generation, data analysis, document creation, and system administration.
Each task is paired with at least one skill sourced from mainstream skill sources: Anthropic-skills~\cite{anthropic-skills-repo}, OpenClaw skills~\cite{openclaw}, and skills.sh~\cite{skillssh}.
We employ an automated evaluation suite comprising static analysis and LLM-based assessment against predefined criteria~\cite{zhu2024swebench, zheng2023llmasjudge}.

\heading{Models.}
We evaluate \sys across eight models spanning three capability tiers to comprehensively assess its effectiveness across varying model capacities.
\emph{SOTA}: claude-opus-4.6~\cite{claude4_6_opus}, deepseek-v3.2~\cite{deepseek2025v3}.
\emph{Mid-tier}: gemini-3-flash~\cite{gemini_flash}, qwen3.5-397b~\cite{qwen3_5_397b}, qwen3.5-122b~\cite{qwen3_5_122b}, claude-3.5-haiku~\cite{claude3_5_haiku}.
\emph{Small}: qwen3-30b~\cite{qwen3_30b}, devstral-small~\cite{devstral_small_2507}.

\heading{Baselines.}
We compare \sys against three baselines.
\emph{No Skill}: the agent executes without any skill loaded.
\emph{Original}: the original skill.
\emph{Skill-Creator}: the skill optimized by Anthropic's Skill-Creator~\cite{anthropic-skill-creator} using claude-opus-4.6.

\heading{Harnesses.}
We select three different open-source agent harnesses to evaluate \sys performance across diverse execution environments.
Each harness offers distinct functionality:
\emph{BareAgent}: a minimal harness that injects the compiled skill
directly into the system prompt without additional abstraction layers.
\emph{OpenCode}~\cite{opencode}: a full-featured code agent harness with batch tool
dispatch capabilities and a rich tool set for code execution.
\emph{OpenClaw}~\cite{openclaw}: a general-purpose agent harness with integrated functional
modules designed to support complex multi-step task execution.

\heading{Methodology.}
We first profile the primitive capabilities of different models, and then apply \sys's AOT and JIT compilation to generate optimized skill variants tailored to each model.
To support the compiled skill artifacts, we extend the agent harnesses with custom loading and management mechanisms that enable \sys's optimizations.
For each task, we generate five diverse input instances and measure the average task completion rate, token consumption, and end-to-end execution latency.
All experiments are conducted on a Mac Mini M4 with 16GB memory.

\begin{figure}[t]
  \setlength{\abovecaptionskip}{3pt}
  \setlength{\belowcaptionskip}{-5pt}
  \includegraphics[width=\columnwidth]{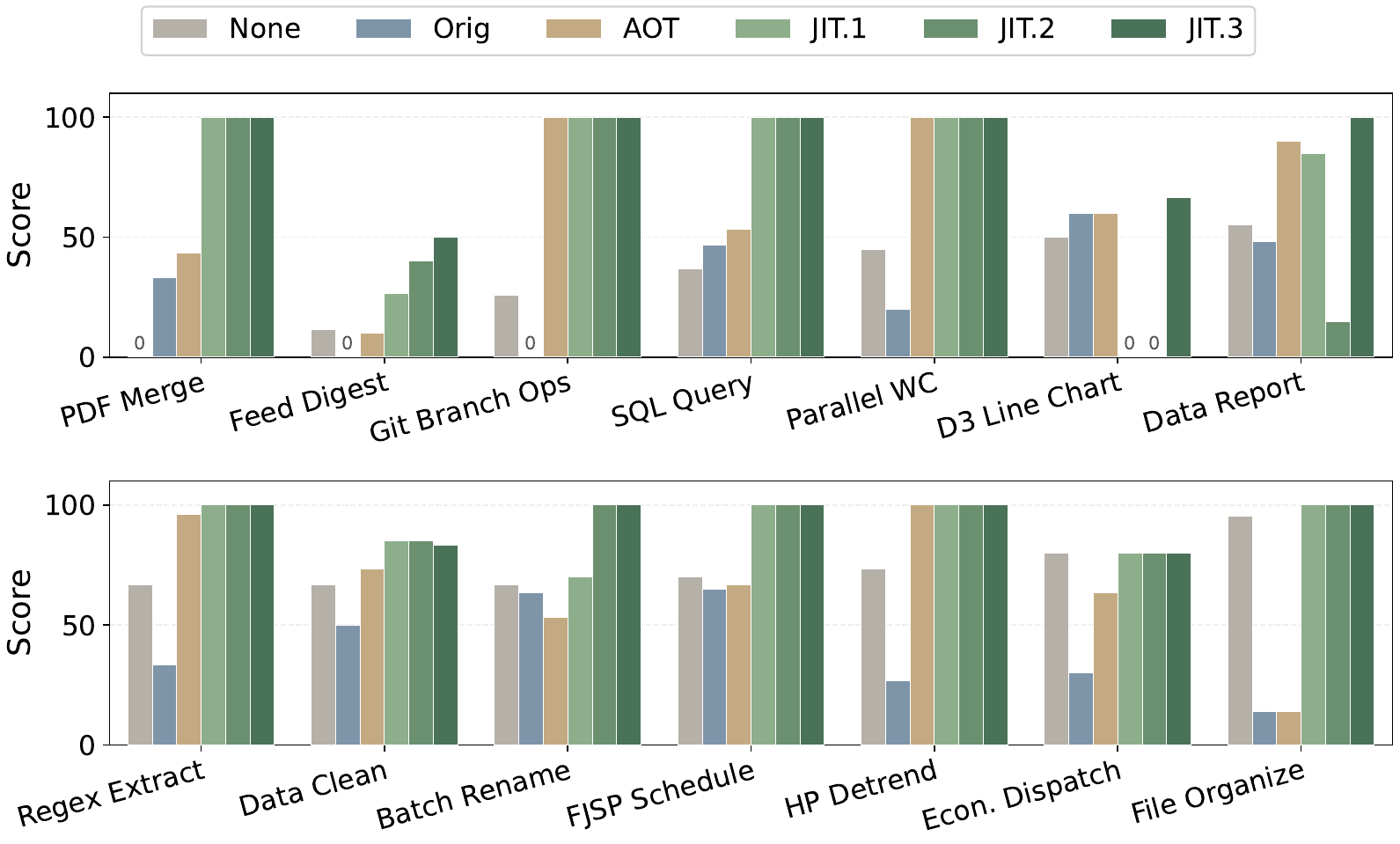}
  \caption{Staged optimization breakdown across 14 skill categories.
  Each group shows scores at six stages: no skill, original skill,
  AOT-compiled skill, and skills optimized in three JIT rounds. }
  \label{fig:e6}
\end{figure}

\subsection{Skill Compilation Effectiveness}
\label{sec:eval:e1}

We first evaluate \sys's effectiveness in improving task completion rates.
We compare \sys-optimized skills against unoptimized baseline skills across different tasks, models, and agent harnesses.
For \sys, the primary gains stem from capability-based compilation and JIT optimization.
Figure~\ref{fig:e1_heatmap} presents compilation effectiveness across eight models and three harnesses.
Cell values represent task completion rates after \sys optimization, while cell color indicates the score improvement (green) or regression (red) relative to unoptimized baseline skills.

Figure~\ref{fig:e1_heatmap} demonstrates \sys's optimization effectiveness (compared with original skills) across diverse models and agent harnesses.
Weaker models benefit more substantially from \sys optimization.
This finding indicates that for most agent tasks, weaker models possess sufficient capability for the logical components but lack proficiency in non-logical aspects, such as generating complex JSON structures or managing environment dependencies.
\sys's compilation optimization addresses these gaps through primitive capability refinement, 
thereby improving overall task completion rates.
For stronger models, improvements are more modest, as these models already handle tasks competently; 
\sys optimization primarily reduces token consumption and execution latency.
Figure~\ref{fig:e1_overview} aggregates average task scores across all four skill variants
for each model and harness.
\sys-optimized skills achieve the highest scores on every model--harness combination.
The improvement over the Skill-Creator baseline grows with decreasing model capability:
on BareAgent, \sys-optimized skills improve over Skill-Creator by 25\% for qwen3-30b and 10\% for devstral-small.
In terms of regressions, \sys-compiled skills reduce task completion rates on only 4.5\% of tasks, compared with 15\% for the original skills.
Using the original skill, OpenCode and OpenClaw differ by up to 13 points,
depending on the model, which shows that harness behavior alone materially changes outcomes.
\sys optimization raises scores on both harnesses and reduces this cross-harness gap to at most 5 points.

\subsection{Staged Optimization Breakdown}
\label{sec:eval:e6}

Figure~\ref{fig:e6} breaks down the contribution of each optimization stage
across 14 skill categories on Qwen3-30B with BareAgent.
Each group shows scores at six stages: no skill, original skill,
AOT-compiled skill, and up to three rounds of JIT optimization.
Tasks used in each round are of the same type but differ in contents.

In 11 of 14 categories, the original skill performs worse than using no skill.
After \sys's AOT compilation, the average task score improves by 88\%.
JIT optimization identifies additional capability defects that were not exposed during AOT compilation,
further improving performance.
After one JIT round, 8 of 14 skills achieve full scores on the
evaluation tasks, and after three rounds, 10 of 14 do so.

JIT optimization can introduce regressions.
For example, in the Line Chart task, the JIT compiler added an example file to the skill,
but the file was too long and triggered parsing errors, which reduced the score.
After three rounds of revision guided by prior failure logs, the JIT compiler eventually fixed this issue.

\subsection{Token and Cost Efficiency with Compilation}
\label{sec:eval:e2}

\begin{figure}[t]
  \centering
  \setlength{\abovecaptionskip}{5pt}
  \setlength{\belowcaptionskip}{-3pt}
  \includegraphics[width=\columnwidth]{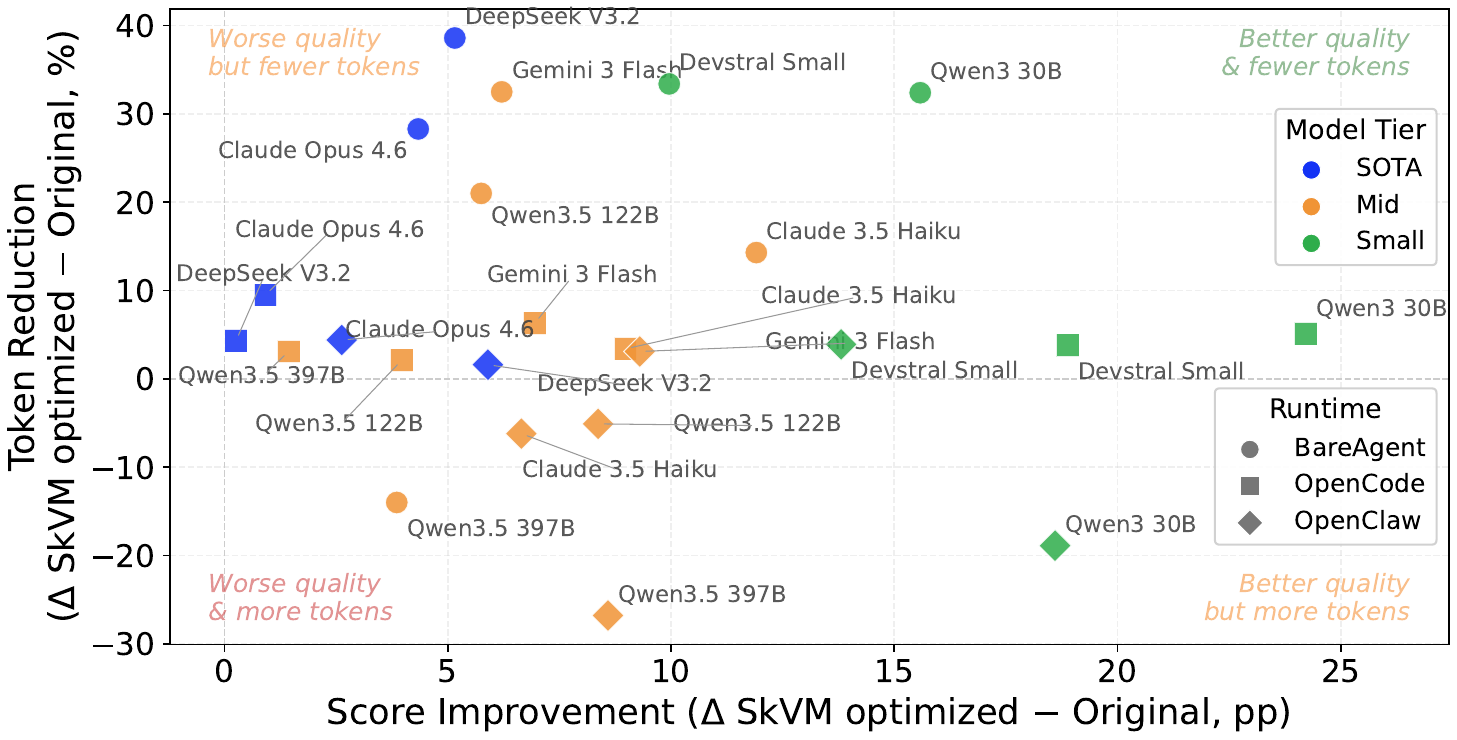}
  \caption{Score gain vs.\ token savings for \sys
    optimization over the Original baseline. Each point is a
    (model, harness) pair. Color = model tier, shape = harness.
    Most points land in the upper-right quadrant: better
    quality \emph{and} fewer tokens.}
  \label{fig:e2_scatter}
\end{figure}

Beyond improving task success rates, \sys provides substantial cost benefits for SOTA LLMs by reducing token consumption.
LLMs can correct execution errors through iterative agent loops, refining actions based on environment feedback.
Therefore, even when tasks eventually succeed, this approach may incur substantial token waste.
Figure~\ref{fig:e2_scatter} shows how task success rates and token costs vary across different model tiers and harnesses after \sys compilation.

Our evaluation reveals that for most models and harness combinations, \sys compilation simultaneously improves task success and reduces token consumption.
For the strongest model and weakest harness pairing (DS-v3.2 + BareAgent), we observe token savings approaching 40\%.
This improvement arises because agent loops incur interaction overhead when correcting environment-related and tool-invocation errors; multiple retries are often required.
By employing JIT compilation, \sys enables models to avoid these error paths entirely, eliminating redundant interactions and saving substantial tokens.

For weaker models, \sys may occasionally consume additional tokens.
This occurs because \sys improves task completion rates, leading to longer agent execution traces compared to the baseline skill, resulting in modest token increases in edge cases.

\begin{figure}[t]
  \centering
  \setlength{\abovecaptionskip}{5pt}
  \setlength{\belowcaptionskip}{0pt}
  \includegraphics[width=\columnwidth]{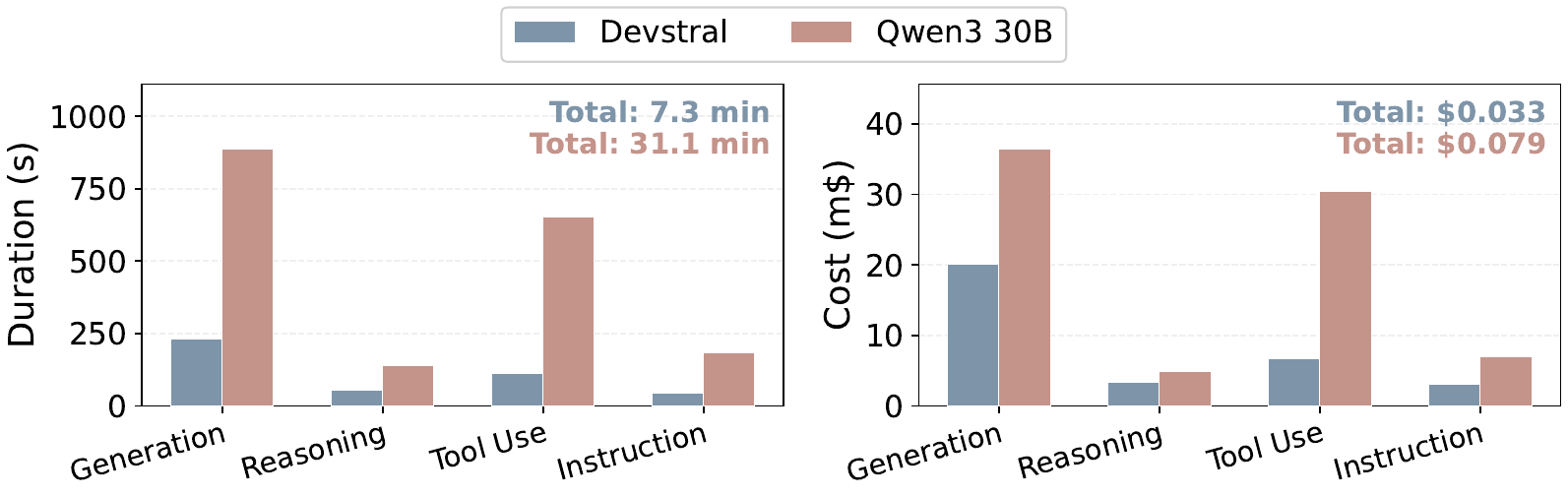}
  \caption{Per-category breakdown of capability profiling
    overhead for two small models. Left: duration in seconds.
    Right: cost in millidollars.}
  \label{fig:profiling_cost}
\end{figure}

\begin{figure}[t]
\centering
\setlength{\abovecaptionskip}{3pt}
\setlength{\belowcaptionskip}{0pt}
\includegraphics[width=\columnwidth]{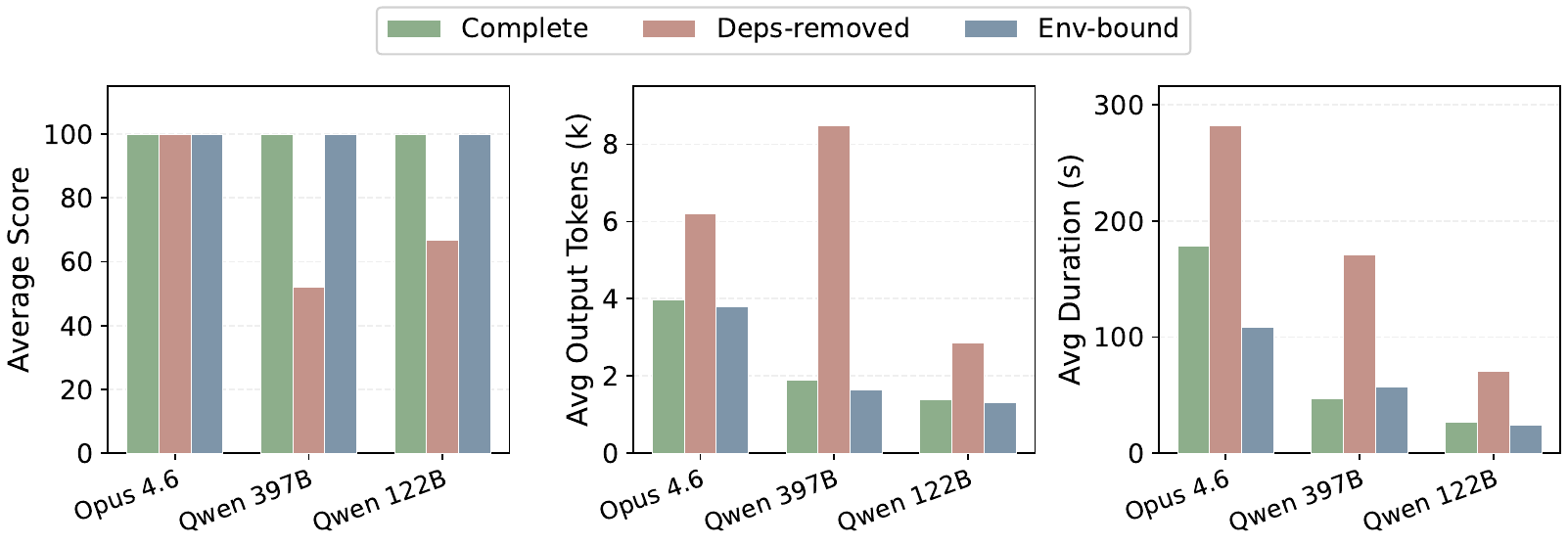}
\caption{Environment binding restores correctness, token
  efficiency, and execution speed. From left to right: average
  score, average token use, and average duration across two
  tasks per model. Env-bound performance returns to
  complete-environment levels for all three models.}
\label{fig:e3}
\end{figure}

\begin{figure*}[t]
  \centering
  \setlength{\abovecaptionskip}{3pt}
  \setlength{\belowcaptionskip}{0pt}
  \includegraphics[width=\textwidth]{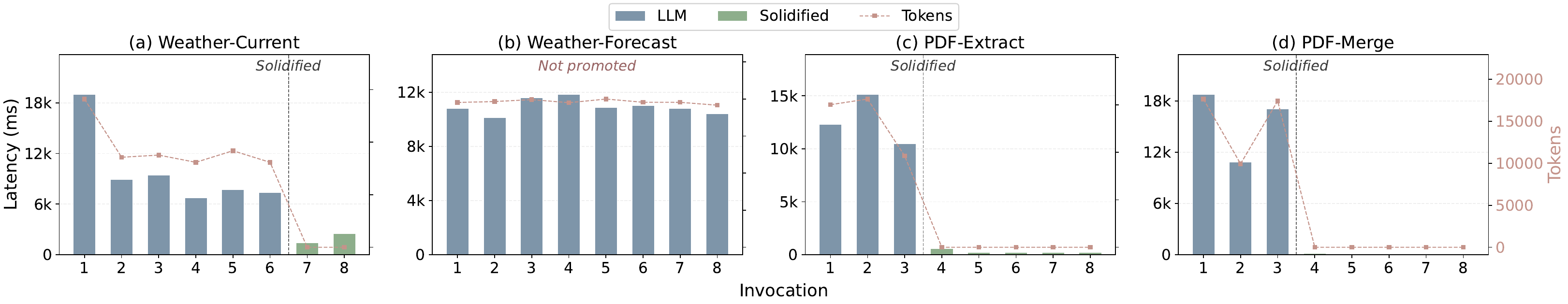}
  \caption{Code solidification across four cases. Blue bars
    show LLM inference latency, green bars show solidified
    execution. The dashed line marks the promotion point.
    Weather-forecast never promotes because the runtime
    model's code patterns diverge from the AOT prediction,
    validating the promotion gate as a safety mechanism.}
  \label{fig:e5_jit}
\end{figure*}

\subsection{Overhead of Target Profiling}
\label{sec:eval:profiling}

Capability-based compilation requires profiling the target once
to build its capability profile (\S\ref{sec:compiler:cap}).
This profiling is a one-time cost per (model, harness) pair,
and the results are cached and reused for all subsequent
skill compilations on the same target.
We measure this overhead on two small models, devstral-small and qwen3-30b. 

Figure~\ref{fig:profiling_cost} shows the per-category breakdown
of profiling duration and cost.
A full profile across all primitive capabilities completes in
7.3 minutes for devstral-small and 31.1 minutes for qwen3-30b,
costing \$0.033 and \$0.079 respectively.

\subsection{Evaluating Environment Binding}
\label{sec:eval:e3}

The preceding two experiments conducted testing in fully provisioned environments. However, in real-world skill scenarios, missing environment dependencies during agent execution are commonplace. Resolving such environmental issues presents significant challenges for weaker LLMs, frequently resulting in skill execution failures. \sys addresses this through an environment binding mechanism in its AOT compilation phase. This mechanism shifts environment dependency installation from the LLM execution phase to pre-execution preprocessing, allowing the LLM to focus solely on skill logic.

Figure~\ref{fig:e3} demonstrates the performance of different models across three environmental configurations: complete, with environment binding, and with missing dependencies.
Stronger models such as claude-opus-4.6 can autonomously recover from missing dependencies, but at the cost of increased token consumption.
For weaker models such as qwen3.5-122b, missing dependencies lead to execution failures.
Environment binding fully restores execution correctness and substantially reduces token consumption.

\subsection{Evaluating Concurrency Extraction}
\label{sec:eval:e4}

\begin{figure}[t]
\centering
\setlength{\abovecaptionskip}{5pt}
\setlength{\belowcaptionskip}{-5pt}
\includegraphics[width=\columnwidth]{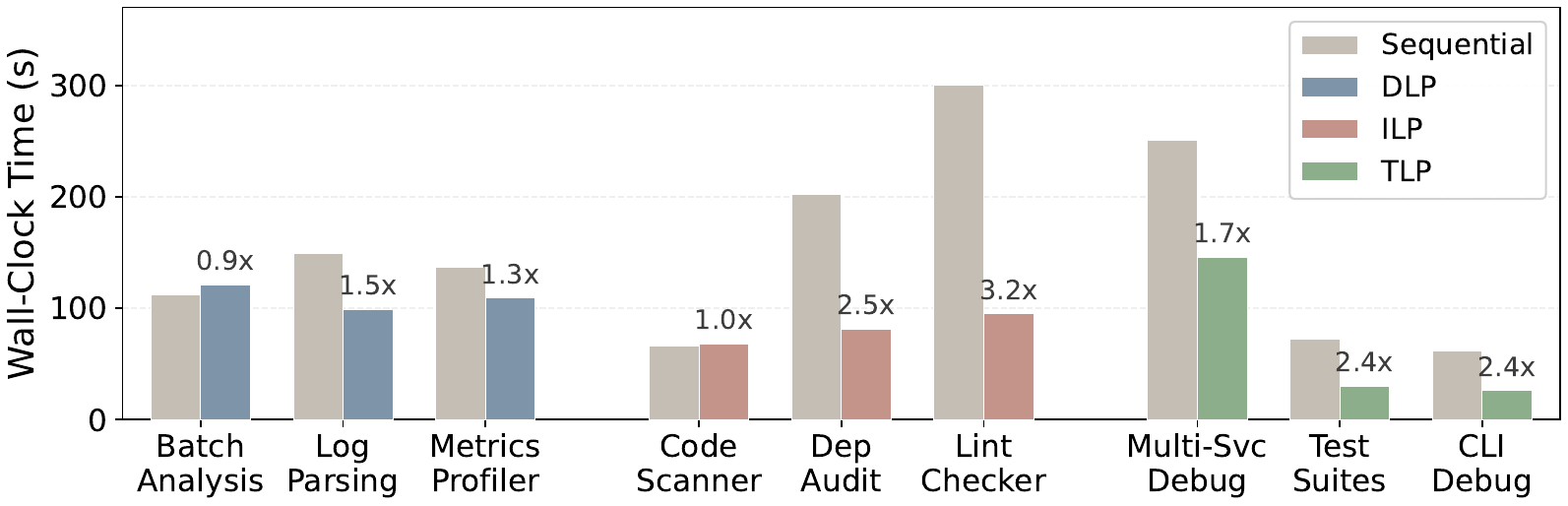}
\caption{Performance of different parallelization strategies across eight tasks.
  Bars show sequential, DLP, ILP, and TLP execution.}
\label{fig:e4}
\end{figure}

\sys extracts more parallelism opportunities within skills and improves efficiency.
We evaluate three parallelism types: data-level parallelism (DLP),
instruction-level parallelism (ILP), and thread-level parallelism (TLP).
DLP applies when instructions within a single skill batch-process large
amounts of independent data. ILP applies when multiple independent instructions
or code segments within a skill can execute concurrently. TLP applies when
multiple independent sub-agents run within a skill, each processing
self-contained subtasks without cross-dependencies. Figure~\ref{fig:e4}
demonstrates \sys's performance improvements across these three parallelism
types.

Experimental results show that \sys's parallelism extraction strategy
achieves up to 3.2$\times$ end-to-end speedup. TLP yields the largest
average improvements because its coarser-grain parallelism produces more
pronounced optimization effects. ILP, operating at the instruction level,
reduces both execution time and the number of LLM invocations. DLP gains
stem purely from data parallelism, such that the performance
improvement scales directly with the degree of data parallelism.
We also observe inherent variability in agent execution due to LLM throughput
fluctuations and varying numbers of agent loop iterations. Consequently,
timing variations within a reasonable range are acceptable.

\subsection{Evaluating Code Solidification}
\label{sec:eval:e5}

We further analyze whether adopting code solidification
mechanisms from JIT optimization can accurately identify
reusable code segments and improve agent execution
efficiency. As shown in Figure~\ref{fig:e5_jit}, we conduct
a detailed analysis on two canonical skills (weather,
document-pdf).
For the two PDF tasks, despite variations in task inputs, the LLM-generated code matches the code signature. Therefore, \sys can apply JIT compilation optimization, directly generating solidified code using the template and input parameters. 
For PDF-extract tasks, JIT optimization reduces execution time from
10,469--15,116\,ms to 206--568\,ms, achieving a 19--50$\times$ speedup.
For the two weather tasks, weather-current requires external weather API calls,
the speedup is bounded by network latency, reducing execution time from 9,000\,ms
to 2,000\,ms, yielding a 5$\times$ improvement. 
In contrast, weather-forecast uses more flexible formats,
so the generated code structures diverge from the code signature and fail to match.
Consequently, all eight invocations rely on
LLM generation to avoid invoking incorrect code.

If solidified code execution causes task failure, \sys's safety mechanism
triggers fallback, re-enabling LLM code generation to ensure correctness.

%% file: discussion.tex
\section{Discussion}
\label{sec:discussion}

\heading{Non-determinism in skill compilation.}
Unlike traditional program compilation, skill compilation takes natural language as input,
which inherently introduces some non-determinism into the compilation process~\cite{wang2023selfconsistency}.
However, the LLM executing skills differs fundamentally from traditional hardware such as CPUs~\cite{hennessy2019architecture},
and possesses an inherent tolerance for input variability that enables skills to execute correctly.
Moreover, \sys's rigorous compilation optimization passes and rollback mechanisms ensure
that compiled skills consistently deliver stable performance improvements across most downstream tasks.

\heading{Capability Coverage.}
The current catalog has 26 primitive capabilities across four categories,
enough to cover the 15,063 skills we analyzed. 
As the skill library grows, we can adopt the iterative approach described in \S\ref{sec:compiler:cap} to expand the primitive capabilities.

\heading{Compilation cost.}
AOT compilation invokes an LLM, which incurs token costs. 
However, since skills are executed repeatedly at runtime,
the compilation overhead is amortized across multiple invocations~\cite{gray1981transaction}.
Furthermore, compiled skills can be shared across users and applications, further reducing the per-skill compilation cost.

%% file: concl.tex
\section{Conclusion}
\label{sec:conclusion}

Skills have emerged as a new code form in the agent era.
However, after analyzing over 100,000 skills, we discover a significant mismatch between skills and underlying LLMs.
To address this, we propose \sys, a compilation and runtime system specifically tailored for skills.
\sys leverages compilation techniques such as AOT and JIT to optimize skill structure, extract environment dependencies, and identify parallelism opportunities.
As models and agent harnesses continue to diversify, skill compilation provides a systematic pathway to transform skills into portable components, rather than fragile prompts.